\documentclass[twocolumn,floatfix]{revtex4}

\usepackage{graphicx,bm,psfrag, amsmath, color}

\usepackage[dvipsnames]{xcolor}

\newcommand{\beq}{\begin{equation}}
\newcommand{\eeq}{\end{equation}}
\newcommand{\beqa}{\begin{eqnarray}}
\newcommand{\eeqa}{\end{eqnarray}}
\newcommand{\ba}{\begin{array}}
\newcommand{\ea}{\end{array}}

\usepackage{youngtab}

\newcommand{\be}{\begin{equation}}
\newcommand{\ee}{\end{equation}}
\newcommand{\bea}{\begin{eqnarray}}
\newcommand{\eea}{\end{eqnarray}}

\newcommand{\Tc}{T_{\mathrm c}}
\newcommand{\Tosc}{T_{\mathrm osc}}
\newcommand{\Tr}{T_{\mathrm r}}
\newcommand{\TSPH}{T_{\mathrm{sph}}}
\newcommand{\TEW}{T_{\mathrm{ew}}}
\newcommand{\Lc}{\Lambda_{\mathrm c}}
\newcommand{\fc}{f_{c}}
\newcommand{\mpl}{m_{\mathrm{pl}}}

\newcommand{\fw}{f_\mathrm{w}}

\usepackage[english]{babel}
\usepackage[utf8x]{inputenc}
\usepackage[T1]{fontenc}

\usepackage[colorinlistoftodos]{todonotes}
\usepackage[colorlinks=true, allcolors=blue]{hyperref}

\begin{document}

\widetext

\begin{flushright}
IPPP/18/93 \\
\end{flushright}

\title{
Out-of-the-box Baryogenesis During Relaxation
}
\author{S.A.~Abel, R.S.~Gupta and J.~Scholtz}
\affiliation{Institute for Particle Physics Phenomenology,
Durham University, South Road, Durham, DH1 3LE}

\date{\today}

\begin{abstract}
\noindent We show that spontaneous baryogenesis occurs automatically in relaxion models if the reheating temperature is larger than the weak scale, provided the Standard Model fields are charged under the U(1) of which the relaxion is a pseudo-Nambu-Goldstone boson. During the slow roll, the relaxion breaks CPT, biasing the thermal equilibrium in favor of baryons, with sphalerons providing the necessary baryon number violation. We calculate the resulting baryon asymmetry, explore the possible constraints on this scheme and show that there is a swath of parameter space in which the current observations are matched. Successful baryogenesis can be achieved for a range of relaxion masses between  $10^{-10}$ and $10^{-5}$ eV. The mechanism operates precisely in the region of parameter space where recent work has shown   relaxion oscillations to be a dark matter candidate.
\end{abstract}

\maketitle

\section{Introduction}

An interesting explicit scenario has been suggested \cite{Graham:2015cka} to explain why the Higgs mass could naturally be much smaller than the fundamental scale. Adapting a 
long-standing idea of Abbott that attempted (unsuccessfully) to explain the smallness of the cosmological constant \cite{Abbott:1984qf}, the {\it relaxion} mechanism incorporates an ingenious interplay of two explicit/anomalous breakings of a Goldstone shift symmetry, to relax the Higgs mass dynamically to values close to the weak-scale. 
Initially the smaller breaking drives the pseudo-Nambu-Goldstone mode  (PNGB), the so-called relaxion, which samples over Higgs masses,  while the 
larger breaking comes in the form of a periodic axion-like potential  
that is proportional to the Higgs mass itself: hence the dynamical evolution of the relaxion stops as soon as the Higgs mass turns on. With a  suitable choice of parameters 
the resultant Higgs mass can be of the correct order.

The central achievement of \cite{Graham:2015cka} was to relate the Higgs mass to a small technically natural parameter, overcoming the 
lack of any direct equivalent to the chiral symmetry breaking that protects quark masses.
The proposal can be made natural in the colloquial sense as well, by adapting clockwork-like scenarios \cite{Kaplan:2015fuy}. 
  
This paper is motivated by the fact that, being a PNGB, the rolling of the relaxion represents a spontaneous CPT violation. Successful baryogenesis scenarios based on such dynamical evolution were proposed in  \cite{Cohen:1987vi,Cohen:1988kt,Cohen:1990py,Cohen:1991iu,Nelson:1991ab,Abel:1992za,Cohen:1993nk} (also see \cite{DeSimone:2016ofp} for a recent review),  and were dubbed spontaneous baryogenesis (SBG). One is naturally led to ask if spontaneous baryogenesis can be incorporated into the relaxion mechanism. 

We will demonstrate that in fact  SBG is virtually generic. It requires only the most minimal augmentation of the relaxion mechanism, namely the addition of a single term to the Lagrangian:
 \begin{equation}
 \label{eq:term}
\mathcal{O}_1 ~=~ \frac{1}{f} \partial_\mu \phi J^\mu ~,
\end{equation}
where $\phi$ is the relaxion and $J^\mu$ is a current of matter fermions that has a component orthogonal to both electromagnetic charge $Q$, and  $B-L$. Moreover we will see that the term in Eq.~\eqref{eq:term} can be generated simply by charging the matter fields under the $U(1)$ corresponding to the relaxion shift symmetry.

Although the term in Eq.~\eqref{eq:term} is invariant, a homogenous time-derivative of $\phi$ amounts to the aforementioned spontaneous violation of CPT,
and it implies an effective chemical potential for baryon number  \cite{DeSimone:2016ofp}.
As described in \cite{Cohen:1987vi,Cohen:1988kt,Cohen:1990py,Cohen:1991iu,Nelson:1991ab} this  biases sphaleron transitions for temperatures around the electroweak  scale where they are active and dominant. In the context of this discussion the time-dependence is a function of the relaxion dynamics itself,  which also ultimately quenches $B+L$ violation while there is still a nett 
baryon asymmetry. Thus remarkably all the required ingredients for SBG are {\it already present} in the relaxion mechanism, provided the term in Eq.~\eqref{eq:term} is generated with sufficient strength. 

\begin{figure}
\noindent \begin{centering}
\includegraphics[scale=0.4, bb=0bp 0bp 6100bp 390bp,clip]{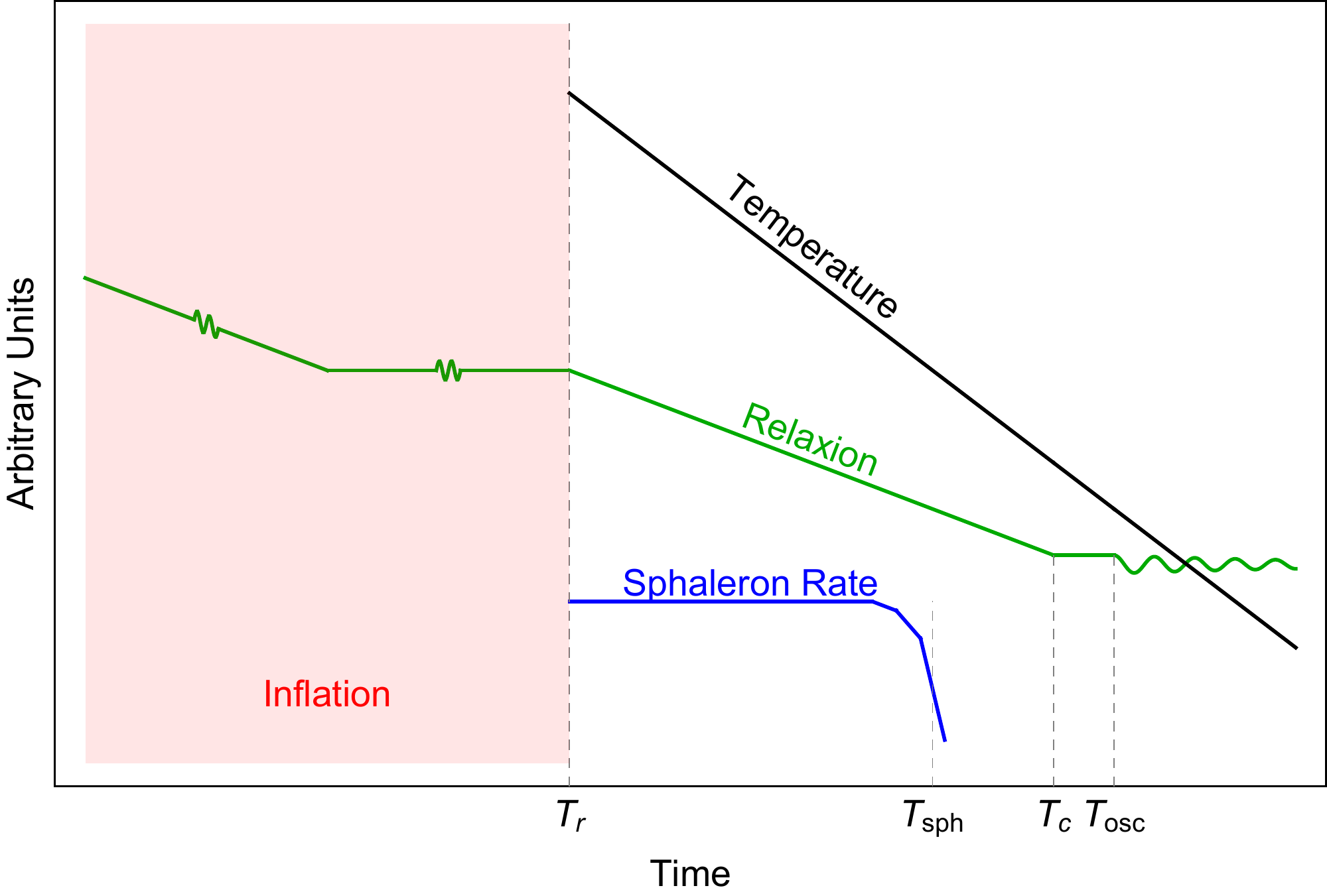}
\par\end{centering}
\protect\caption{The pattern of evolution required for successful baryogenesis during cosmological relaxation. \label{fig:sbg}}
\end{figure}

Successful SBG occurs if the cosmological evolution follows the pattern shown in fig.~\ref{fig:sbg}. There is a period of inflation (red) and reheat to temperature $\Tr$ (black). After reheat and for a period until the temperature drops below $\TSPH \sim \TEW \sim 130$ GeV \cite{DOnofrio:2014rug}, $B+L$ violating transitions (blue) are active in the plasma, dying away exponentially fast below $\TSPH$. 
During this period the relaxion evolves homogeneously (green) towards its final value which it reaches at temperature $\Tc$. Thus one has to satisfy the constraint    $\Tr > \TSPH > \Tc$, which 
as we shall see can be satisfied without violating any bounds.
Subsequently as the temperature drops further the relaxion starts to oscillate about the local minimum and becomes a dark matter candidate~\cite{Banerjee:2018xmn}. It is striking that  the baryon asymmetry and the dark-matter matter relic density turn out to be correct in roughly the same region of parameter space.

\section{Description of the Mechanism}

\subsection{The Relaxion}
\label{sec:relaxion}

Let us begin by describing in more detail the cosmological relaxation  mechanism itself, proposed in Ref.~\cite{Graham:2015cka}. Here we focus on models with Hubble friction; for alternative models involving particle production see Ref.~\cite{Hook:2016mqo, Fonseca:2018xzp} and Ref.~\cite{Son:2018avk} where the latter also implements leptogenesis in this scenario. The defining feature of relaxion models is that the Higgs mass squared term, $\mu^2$,  is dynamical and depends on the classical value of the slowly rolling relaxion, $\phi$. We will consider the following typical potential,
\begin{equation}
V_{\rm roll}~=~(g  \phi-M^2   +\cdots) H^\dagger H +\lambda (H^\dagger H)^2+  g M^2 \phi +\cdots ,
\end{equation}
where $M$ is a cut-off scale. Suppose that initially the relaxion field has a value $\phi\gtrsim M^2/g$ such that $\mu^2>0$, and electroweak symmetry is unbroken. During inflation, the relaxion field rolls due to the linear term, and in the process scans the value of $\mu^2$.  As the relaxion field crosses the point $\phi = M^2/g$ ,  $\mu^2$ becomes negative triggering electroweak symmetry breaking. At this point,  the Higgs gets a vacuum expectation value (VEV), turning  on a crucial ingredient of relaxion models, the so-called backreaction potential, 
\begin{eqnarray}
V_{\rm br}~=~ \Lc^4 \cos \frac{\phi}{\fw}~=~ \tilde{m}^{4-j} \langle H\rangle^{j} \cos \frac{\phi}{\fw}~.
\end{eqnarray}
Here $j<4$ is a model-dependent integer. We will describe how such a potential is generated below, but for the moment we note that the important feature is its periodicity, which causes the relaxion comes to a halt at $\phi=\phi_0$ satisfying, 
\begin{eqnarray}
V'_{\rm roll}(\phi)+V'_{\rm br}(\phi)\,=\, 0 ~\Rightarrow~ g M^2 \,=\, \frac{ \Lc^4}{\fw}\sin  \frac{\phi_0}{\fw}.~
\label{eq:vprime}
\end{eqnarray}
Due to quantum spreading, when it stops, the relaxion field is spread across multiple vacua with weak scale Higgs vacuum expectation value. While,  generally the phase  at these minima $\frac{\phi_0}{\fw} \sim 1$,  it can be arbitrarily close to 0 or $\pi/2$ in a small fraction of these vacua  (see for eg. Ref.~\cite{Gupta:2018wif}).  A large hierarchy between the Higgs VEV and the cutoff scale can therefore be achieved if $g$ is very small. This smallness of $g$ is radiatively stable, because one recovers the discrete symmetry $\phi\to 2 \pi n \fw, n\in Z$, in the limit $g\to 0$. Note that the relaxion travels a distance $F\sim M^2/g\gg f_w$ in field space which can be problematic  as discussed in  Ref.~\cite{Gupta:2015uea}. This feature can, however,  be incorporated in a consistent way in clockwork models~\cite{Choi:2014rja, Choi:2015fiu, Kaplan:2015fuy} where the $g$-dependant parts of the potential arise from  periodic terms with the larger  periodicity $2 \pi F$.

Let us briefly now discuss how the backreaction contribution can arise by describing the non-QCD model of Ref.~\cite{Graham:2015cka}. In this model, $\phi$ is the axion of a new strong sector, and the Higgs couples to two vectorlike leptons charged under the strong group, as follows:
  \begin{eqnarray}
  {\cal L}~=~ y_1 LHN+y_2 L^cH^\dagger N^c -m_L L L^c-m_N N N^c+h.c.
  \nonumber
  \end{eqnarray}
Here $(L, N)$ have the same electroweak quantum numbers as the SM lepton doublet and   right-handed neutrino, respectively, and $(L^c, N^c)$ are in the conjugate representations. 
Integrating out $L, L^c$ below their mass,  we obtain a Higgs contribution to the mass of $N$, $\Delta m_N= \frac{y_1 y_2 \langle H\rangle^2}{m_L}$. 
As long as $m_N\ll 4 \pi \fc \ll m_L$, the fermion $N$ forms an electroweak preserving  condensate, and  the relaxion, being the axion of the new strong group, gets the following contribution to its potential due to non-perturbative effects:
\begin{align}
V_{\rm br} &~\simeq~ - 4 \pi \fc^3 \Delta m_N \cos \frac{\phi}{\fw} \notag\\
& ~=~ -\frac{4 \pi \fc^3 y_1 y_2\langle H\rangle^2 }{m_L}\cos \frac{\phi}{\fw} ~\stackrel{\rm def}{=}~ -\Lc^4 \cos \frac{\phi}{\fw}~.
\label{brstr}
\end{align}
 There are also radiatively generated, Higgs independent, contributions to the potential. To ensure that these are subdominant relative to Eq.~\eqref{brstr}, we  must have $\fc \lesssim v, m_L\lesssim 4 \pi v$ and $\Delta m_N\lesssim 4 \pi \fc$~\cite{Graham:2015cka}. Taken together these conditions give the upper bound,
\begin{equation}
\Lc^4 ~\lesssim~ (16 \pi ^2) v^4~.
\label{eq:quantcorr}
\end{equation}

The period immediately after inflation is when baryogenesis can take place and appears to be somewhat generic. Our main assumption is that the reheat temperature after inflation, $\Tr$, is higher than the electroweak scale, $v$. Since $\fc<v$ one can conclude that  $\Tr> \Tc \sim \sqrt{4 \pi} \fc$, where $\Tc$ is the critical temperature corresponding to the chiral phase transition  of the new strong group. Note that a separation of scales $\Lc \ll \fc\sim \Tc$ is technically natural. Therefore under this assumption the backreaction potential vanishes after inflation, and the relaxion enters a second short period of rolling.  

 As the universe cools and the temperature drops below $\Tc$ again,  the `wiggles' of the backreaction potential reappear.  During the second phase of rolling the relaxion  obeys the attractor solution $V'\simeq  5 H \dot{\phi}$ and   gets trapped again by  the  backreaction potential, provided the temperature induced Hubble friction is always larger than its mass, i.e. $m_\phi\lesssim H \sim T^2/\mpl$~\cite{Kawasaki:2011pd}.  In Refs.~\cite{Banerjee:2018xmn,Choi:2016kke}  a more precise condition for the relaxion to get trapped was derived from a full solution to the equations of motion, namely
 \begin{equation}
m_\phi ~\lesssim~ 5 H(\Tc) ~.
\label{eq:slowroll}
 \end{equation}
This  condition  was obtained by demanding  that the relaxion does not pick up enough kinetic energy to overshoot the barriers once the backreaction potential reappears.
 
 In the second phase of rolling the relaxion is displaced from its original stopping point by a misalignment angle given by~\cite{Banerjee:2018xmn},
  \begin{equation}
\Delta \theta~=~\frac{\Delta \phi}{f} ~\simeq~ \frac{1}{20}\left(\frac{m_\phi}{H(\Tc)}\right)^2 \tan \frac{\phi_0}{f}~.
\label{eq:misalignment}
 \end{equation}
One can check that this changes the Higgs mass  by a very small amount. Once the temperature drops to a value such that $H (\Tosc) < m_\phi/3$, the relaxion starts oscillating about the local minimum. As shown in Ref.~\cite{Banerjee:2018xmn}, these oscillations give rise to a relic abundance given by, 
  \begin{equation}
\Omega h^2 ~\simeq~ 3 \Delta \theta^2 \left(\frac{\Lambda_d}{1~{\rm GeV}}\right)^4  \left(\frac{100~{\rm GeV}}{\Tosc }\right)^3~.
\label{eq:dm}
 \end{equation}

\subsection{Spontaneous Baryogenesis}
\label{sec:spbar}

We now proceed to show that baryogenesis is a generic consequence of the relaxion mechanism discussed in the previous section. We begin with an explanation for the operator in Eq.~\eqref{eq:term}, and then in the following subsection estimate the baryon asymmetry that it generates. 

\subsubsection{Getting $\mathcal{O}_1$ from Yukawa couplings}

The simplest explanation for the operator ${\cal O}_1$ in Eq.~\eqref{eq:term} is that it arises because the SM fermions are charged under the global symmetry for which the relaxion is a PNGB. In this case   the Yukawa couplings of the SM carry non-zero charge under the relaxion shift symmetry. To explore this possibility we will generalize the discussion in Ref.~\cite{Abel:1992za} in order to deal correctly with anomalies. We begin with the Yukawa-induced masses for the fermions during the relaxion evolution, which in the unitary gauge take the form 
\begin{eqnarray}
\label{Yukawa}
 {\cal L}_{m} & ~=~ & -\lambda_{u} v\, [u_{r}^{\dagger}e^{-i\theta_u}u_{l} +
u_{l}^{\dagger}e^{i\theta_u}u_{r}] \nonumber \\
&  & - \lambda_{d}v\,  [d_{r}^{\dagger}e^{-i\theta_d}d_{l} +
d_{l}^{\dagger}e^{i\theta_d}d_{r}] \nonumber \\
&  & - \lambda_{e}v \, [e_{r}^{\dagger}e^{-i\theta_e}e_{l} +
e_{l}^{\dagger}e^{i\theta_e}e_{r}]~,
\end{eqnarray}
where obviously $v=\langle H\rangle$, and a summation over generations 
is implied (and we neglect flavour mixing). The $\theta_i$ are time-dependent phases driven by the relaxion mechanism,
\be 
\label{eq:charges}
\theta_i ~=~ q_{\lambda_i} \frac{\phi}{f}~,
\ee 
 where the $ q_{\lambda_i}$ are the charges of the Yukawa couplings (i.e. the difference in global charge between the respective left and right handed fermion) under this symmetry. The Higgs VEV, $v(t)$, is of course also time dependent, although that has only a minor effect on the baryon asymmetry. 
For simplicity we will for this illustrative example take the phases $\theta_i$ to be generation independent and also neglect flavour changing. 
As shown in Ref.~\cite{Davidi:2018sii}, a more complicated generation dependent structure, also addressing the SM fermion masses and mixings, can be implemented involving a Froggatt-Nielsen mechanism, with $\frac{\phi}{f}$ playing the role of the familon. 

To determine the physical  consequences of the evolving phases in Eq.~\eqref{Yukawa}, one can remove them at the expense of inducing 
baryon and lepton currents. Broadly speaking this must be equivalent to a term of the form in Eq.~\eqref{eq:term} with  
$
J^\mu 
$ being some combination of only right-handed currents. We know this because one is free to remove the phases with a rotation on the 
right-handed fermions only, which avoids topological $SU(2)$ terms.

To see this in more detail, first we remove the phases in Eq.~(\ref{Yukawa}) by making the rotations,
\begin{eqnarray}
\label{eq:rots}
q_l \rightarrow e^{i\omega}q_l &;&
u_r \rightarrow e^{i(\omega-\theta_u)}u_r ~;~
d_r \rightarrow e^{i(\omega-\theta_d)}d_r ~; \nonumber\\
\ell \rightarrow e^{i{\overline\omega}}\ell  &;&
e_r \rightarrow e^{i({\overline\omega}-\theta_e)}e_r ~,
\end{eqnarray}
where we are allowed two arbitrary phases, $\omega$ and $\overline
\omega$ (corresponding to $B$ and $L$ rotations), upon which of course the eventual physics should not depend. 

The corresponding change in the classical action is,
\begin{align}
\label{ds00}
\delta S_0  ~=~  & -\int {\rm d}^4 x \left[
{\overline q}_l \gamma^\mu q_l\partial_\mu \omega +
{\overline u}_r \gamma^\mu u_r\partial_\mu (\omega-\theta_u)
\right.
\nonumber \\ 
& +
{\overline d}_r \gamma^\mu d_r\partial_\mu (\omega-\theta_d) +
{\overline \ell} \gamma^\mu \ell \,\partial_\mu {\overline\omega}
\nonumber \\ 
 & 
 + \left. {\overline e}_r \gamma^\mu e_r\partial_\mu ({\overline\omega}-\theta_e) +\ldots\right]~,
\end{align}
where the ellipsis refers to the accompanying shift in the mass terms that removed the phases. At the same time the rotation in Eq.~\eqref{eq:rots}  induces contributions from 
global anomalies and finite temperature triangle diagrams, which combined (and neglecting all masses except $m_{t,b}$) are 
\begin{align}
\label{ds02}
 \delta S_1  ~=~ &
\int {\rm d}^4 x 
 \left[ \left(18 \omega + 6{\overline\omega} -  \frac{3}{2}\Lambda_t \theta_u- \frac{3}{2}\Lambda_b \theta_d\right)
\frac{g_2^2 F_2 {\tilde F}_2}{32 \pi^2} \right.  \nonumber \\ 
 \mbox{\hspace{-0.01cm} $-$ }
  \mbox{\hspace{-0.05cm}\huge $($\hspace{-0.13cm}}\,
   \frac{9\omega}{2} & \hspace{-0.07cm}+\hspace{-0.07cm} \frac{3{\overline\omega}}{2} \hspace{-0.04cm}
- \hspace{-0.04cm} \mbox{\small ${(\mbox{\normalsize 4}-\frac{17}{24}\Lambda_t) }$}\theta_u \hspace{-0.04cm}
- \hspace{-0.04cm} \mbox{\small{$ (\mbox{\normalsize 1}-\frac{5}{24}\Lambda_b) $}} \theta_d \hspace{-0.04cm}
- \hspace{-0.04cm} 3 \theta_e
 \mbox{ \hspace{-0.2cm}\huge $)$ \hspace{-0.2cm}}
 \frac{g_Y^2 F_Y {\tilde F}_Y}{32 \pi^2} ~~~~~  \nonumber \\ 
  &  ~+ 
\left.   \mbox{\huge $($}\,
 (3 - \Lambda_t) \theta_u + (3 - \Lambda_b) \theta_d   \mbox{\huge $)$}
\frac{g_3^2 G {\tilde G} }{32 \pi^2} \right]  ~,
\end{align}
where the finite temperature pieces are given by
\be
\Lambda (m,T) ~=~ \sum_{n\in { Z}}  \frac{8\pi T m^2 }{3 (m^2 + (2n+1)^2 \pi^2 T^2 )^{3/2}}  ~.
\ee
At high temperatures, $T\gg m$, these terms can be approximated as  
\begin{equation}
\label{lambda}
\Lambda_{T\gg m} ~=~ \frac{14}{3}\zeta(3)
\frac{m^2}{\pi^2 T^2}
\left( 1-
\frac{93 \zeta(5)}{56 \zeta(3)}\frac{m^2}{\pi^2 T^2}
+\ldots \right)~.
\end{equation}
Thus we have implicitly already taken the infinite $T$ limit for all but the top and bottom masses, by setting their $\Lambda$'s to zero. In the small $T$ limit, one instead finds 
\begin{equation}
\label{lambda2}
\Lambda_{T\ll  m} ~=~ \frac{8}{3} ~.
\end{equation}

Here the requirements for baryogenesis during the relaxion mechanism start to diverge from those of electroweak baryogenesis in Ref.~\cite{Abel:1992za}.
In  particular the anomalous $G\tilde{G}$ terms for the strong coupling should be zero in the final vacuum such that an ${\cal O}(1)$ strong CP phase is not generated once the relaxion stops. This motivates   (at least for generation independent phases) models with
the particular  physical choice, 
\be 
\theta_d~=~ - \, \theta_u~.
\label{eq:afree}
\ee
Under this assumption
we may neglect the effect of strong sphalerons \cite{McLerran:1990de,Giudice:1993bb}. 
We now follow \cite{Abel:1992za} and remove the $SU(2)$ topogical piece in the action as well, by setting 
\begin{align}
{\omega} ~~&=~ \frac{1}{12} (\Lambda_t-\Lambda_b) \theta_u - {1\over 3} \overline{\omega} ~~\approx  ~-\,{1\over 3} \overline{\omega} ~.
\end{align}
This particular choice of $ \omega$  
puts all the important physics in the $\delta S_0$ term:
\begin{align}
\delta S_0  ~\approx~ &  \int {\rm d}^4 x \, \left[ {\overline u}_r \gamma^\mu u_r \partial_\mu \theta_u - {\overline d}_r \gamma^\mu d_r\partial_\mu  \theta_u 
  \right. \nonumber \\
&    ~~~~~~~~~~~~~ + \left. {\overline e}_r \gamma^\mu e_r\partial_\mu \theta_e 
  + J^\mu _{B-L} \partial_\mu {\overline \omega}~ \right]  ~.
  \label{ds002}
\end{align}
Note that, since we will ultimately set $B-L=0$, the arbitrary phase $\overline \omega $ can have no effect on the physics. 
Inserting Eq.~\eqref{eq:charges}, the remaining part can as promised be interpreted as a term of the form in Eq.~\eqref{eq:term}, with  
\begin{equation} 
\label{currents}
J^\mu ~=~ q_{\lambda_u} (J^\mu_{u_r} - J^\mu_{d_r}) + q_{\lambda_e} J^\mu_{d_r} ~.    
\end{equation}

It is straightforward to generalise the above discussion to generation dependent charges if we keep in mind some crucial distinctions. First of all, for general charges it is not possible to remove the phases from the Yukawa terms by redefinitions of the singlet quarks alone.  The redefinition of the doublet quarks  will then  generate a $\phi F_2 \tilde{F}_2$ term that can be removed by a  $B+L$ rotation. Finally the  condition in Eq.~\eqref{eq:afree} for not generating a strong CP phase  is generalised to the condition   that the global symmetry  has no triangle anomaly with QCD. Such an anomaly free charge assignment that, at the same time, explains the SM fermion masses and mixings was presented in Ref.~\cite{Davidi:2018sii}

\subsubsection{Getting Baryon Asymmetry from $\mathcal{O}_1$}

\noindent Let us now see how the  operator  $\mathcal{O}_1$ with a current generically of the form, 
\begin{equation}
J^\mu ~=~ \sum_i q_{i} J^\mu_i  + \mathrm{other\; spin\; particles}~,
\end{equation} 
where $i$ goes over all the spin 1/2 particles, leads to a baryon asymmetry.   We can continue with the approach of \cite{Cohen:1991iu,Nelson:1991ab,Abel:1992za}, noting that 
in the presence of the operator $\mathcal{O}_1$ the background value of $\dot{\phi}$ shifts the energy of particles and antiparticles differently (implying a spontaneous breaking of CPT symmetry). This is equivalent to
\begin{align}
\mu_i &= - \bar{\mu}_i ~=~ 
 q_{i} \dot{\phi}/f + (B_i-L_i) \mu_{B-L} + Q_i \mu_{Q}~,
\end{align}
where $\mu_i$ and  $\bar{\mu}_i$ are the chemical potentials for particles of species $i$ and their anti-particles respectively,   $Q_{i}$ is the electromagnetic charge and $\mu_{Q,B-L}$ are Lagrange multipliers introduced to enforce conservation of $Q$ and $B-L$. As a result, as long as there is a source of $B+L$ violation, the following  asymmetry in the 
species $i$ is generated:
\begin{align}
\label{eq:rho}
n_{i} - \bar{n}_i &~=~  f(T,\mu) - f(T,\bar{\mu})\\
&~=~ g_i \left( q_{i}\dot{\phi}/f + (B_i-L_i) \mu_{B-L} +Q_i \mu_{Q}\right) \frac{T^2}{6}~, \nonumber
\end{align}
where we assumed $\mu \ll T$ and the factor $g_i$ incorporates colour, as well as two spin degrees of freedom for each Weyl fermion, a factor of two for complex scalars, and three 
for massive gauge bosons. The chemical potentials $\mu_Q$ and $\mu_{B-L}$ are then  
determined by solving $n_{Q} = n _{B-L} =0$.

For the case at hand, treating 
 ${\dot \theta}_i=q_{\lambda_i} \dot{\phi}$ as classical homogeneous background fields, with the $q_i$'s taken from Eq.~\eqref{currents},  we find 
\begin{align}
\mu_{B-L} &  ~=~  -\dot{\overline \omega}-{ 
{ 3 \left( (7+n) \, \dot\theta_e   
+12 \, \dot\theta_u \right)}\over 
{111+13 n} } ~,\nonumber\\
\mu_{em} & ~=~  \frac{3}{2} 
{{ (5 \, \dot\theta_e   
-39 \, \dot\theta_u )}\over 
{111+13 n} }~,
\end{align}
where in order to keep contact with  \cite{Cohen:1991iu,Nelson:1991ab,Abel:1992za} we have allowed for $n$ charged Higgs scalars with charge $\pm1$.

Finally substituting into Eq.~\eqref{eq:rho} gives 
\begin{align}
n_{B}\ = \ n_{L} & ~=~  \frac{3}{(111+13 n)}\left(
(33+4 n) \dot\theta_e+9\dot\theta_u\right)
\frac{T^2}{6}~,\nonumber \\
& ~=~  
g_{\rm SB}
\frac{\dot\phi}{F}\frac{T^2}{6}~,
\end{align}
where \be
\label{gsb}  g_{\rm SB}  ~=~  
\frac{ 3\left( (33+4 n) q_{\lambda_e}+9 q_{\lambda_u}\right)}{(111+13 n)} ~
\ee  
is a constant of order one.

Note the physical interpretation which is clear from Eq.~\eqref{eq:rho}: there is a bias for $J^\mu$ production in the plasma, but $J^\mu$ does not preserve $B-L$ or $Q_{}$.
Therefore 
any production of $J^\mu$ must be accompanied by a
compensating production of $J_{B-L}^\mu$  and $J^\mu_{Q_{}}$. 
As a consequence even $J^\mu = {\overline e}_r \gamma^\mu e_r $, with no baryon content at all, leads to a baryon asymmetry! (One should be mindful of the 
background assumption that left-handed and right-handed fields remain in chemical equilibrium throughout, via the Yukawa couplings.)

As expected, there is no dependence of the answer 
on $\overline\omega$, so the
physics  does not care about how the phase absorption is
arbitrarily divided between left and right handed fields.
As a second check of these expressions, taking $\theta_u = - \theta_d =- \theta_e = -2 \overline {\omega}$ gives a current in Eq.~\eqref{ds002} proportional to hypercharge, 
as it should because in this case the 
Yukawas have relaxion charges proportional to the hypercharges of the Higgses appearing in the couplings: then Eq.~\eqref{gsb} yields  the answer of \cite{Cohen:1991iu,Nelson:1991ab}, 
relevant for Higgs-driven electroweak baryogenesis. (The relaxion charges here are of course completely general.)

To conclude this section therefore, a term equivalent to Eq.~\eqref{eq:term} is generated if the relaxion is entangled with the 
fermion mass structure, and it leads to a net baryon number density of order
\begin{equation}
\label{eta1}
\eta ~\equiv~ \frac{n_B}{s} ~=~ g_{\rm SB}\frac{\dot\phi}{f}\frac{T^2}{6} \times  \frac{45}{2\pi^2 g_* T^3} ~=~ \frac{15}{4\pi^2}\frac{g_{\rm SB}}{g_*} \frac{\dot \phi}{f T}~.
\end{equation}
where the particular linear combination of currents in $J_\mu$ determines the value of $g_{\rm SB}$.

\subsection{Baryogenesis from the relaxion}

Having collected the components, we now proceed to see how they fit together in a generic relaxion scenario \cite{Graham:2015cka}, and derive constraints. Typically the relaxion stops its slow-roll much before  inflation ends (we would need extreme fine-tuning in order to guarantee that the relaxion travels far in field space but stops at a time close to the end of inflation). Therefore the first slow roll of the relaxion is not suitable for baryogenesis {as  any baryon asymmetry generated would be diluted away.}  As discussed above however, there is a generic scenario  in which the reheat temperature $\Tr$ at the end of the inflation is high enough to restore the chiral symmetry breaking that is the source of the oscillatory potential for the relaxion. Consequently the relaxion undergoes a second stage of rolling, this time in a  radiation dominated phase. It is in this secondary period of rolling that  the  time derivative of the relaxion can drive  baryogenesis.

In order for the mechanism to operate, we first need to ensure that there is an operative source of baryon number violation, which here is provided by electroweak sphalerons. Sphaleron transitions must be faster than the Hubble rate for some region above the temperature  $\TSPH$ at which the sphalerons decouple.

Second, in order to break CPT, the relaxion has to couple to a suitable current $J_\mu$  through an operator such as that 
in Eq.~\eqref{eq:term}. The presence of this operator together with a slowly rolling $\phi$ leads to a baryon asymmetry given by Eq.~\eqref{eta1} with $T=\TSPH$. 

 Using Eq.~\eqref{eta1} and $\dot{\phi} \sim V'(\phi)/(5H)$ (see Sec.~\ref{sec:relaxion}), the generated baryon number asymmetry is
\begin{equation}
\eta~=~ \frac{15}{4\pi^2} \frac{g_{\rm SB}}{g_*^{3/2}}\frac{m_\phi^2 \mpl }{ \TSPH^3} \frac{\fw}{f}~,
\label{eq:eta}
\end{equation}
where the experimentally measured value of $\eta = 8.7 \times 10^{-11}$ \cite{Aghanim:2018eyx} and the relaxion mass is
\begin{equation}
    m_\phi^2 ~=~ \Lc^4/\fw^2~.
    \label{eq:mass}
\end{equation}
From the first of these expressions one can conclude that
\begin{equation}
\frac{\fw}{f} ~=~ \eta \left( \frac{15}{4\pi^2} \frac{g_{\rm SB}}{g_*^{3/2}}\frac{m_\phi^2 \mpl }{ \TSPH^3} \right)^{-1} \sim 10^9 \left(\frac{m_\phi}{10^{-5}\;\mathrm{eV}}\right)^{-2}.
\end{equation}
Therefore successful baryogenesis requires a large scale separation between $f$ and $\fw$. This is acceptable because the relaxion dynamics is already driven by the  much larger scale separation, $F/\fw \sim (M/\Lc)^4$ where $F\sim M^2/g$ is  the total field excursion. The hierarchy $f\ll \fw \ll F$ can easily be obtained from the clockwork mechanism~\cite{Choi:2014rja, Choi:2015fiu, Kaplan:2015fuy} with the operator in Eq.~\eqref{eq:term} arising from the first site of the clockwork, the backreaction potential from an intermediate site and the rolling potential from the last site. In such a setup $f$ would be the only fundamental physical scale, while $\fw, F$ would be fictitious scales.

The prospects for baryogenesis are determined by five unknown parameters ($\fw$, $f$, $\fc$, $\Lc$, $m_\phi$) which, due to the two relations in Eqs.~\eqref{eq:eta} and \eqref{eq:mass}, represent three independent variables. Since the sphalerons must decouple before the back-reaction potential re-emerges at temperature $T_c$, we need  $ \TSPH \gtrsim \Tc$, and so we choose $\fc = \TSPH/\sqrt{4\pi}$ to saturate this bound. As we will see, the bounds for smaller $\fc$ can easily be determined by rescaling. Hence we can display the results in the plane of $m_\phi$ and $f$. The other two variables are then related by:
\begin{align}
    \fw &~=~ \left(\frac{\sqrt{8}\pi^2}{9\sqrt{5}} \eta g_*^{3/2}\frac{\TSPH^3 }{ \mpl}\right) \frac{f}{m_\phi^2}~,\\
    \Lc^4 &~=~  \left(\frac{\sqrt{8}\pi^2}{9\sqrt{5}} \eta g_*^{3/2}\frac{\TSPH^3 }{ \mpl}\right)^2 \frac{f^2}{m_\phi^2}~.
\end{align}

Now, for a region in the $m_\phi,f$-plane to successfully produce the baryon asymmetry, it must satisfy the following constraints:
\begin{figure}
\noindent \begin{centering}
\includegraphics[width=0.45\textwidth]{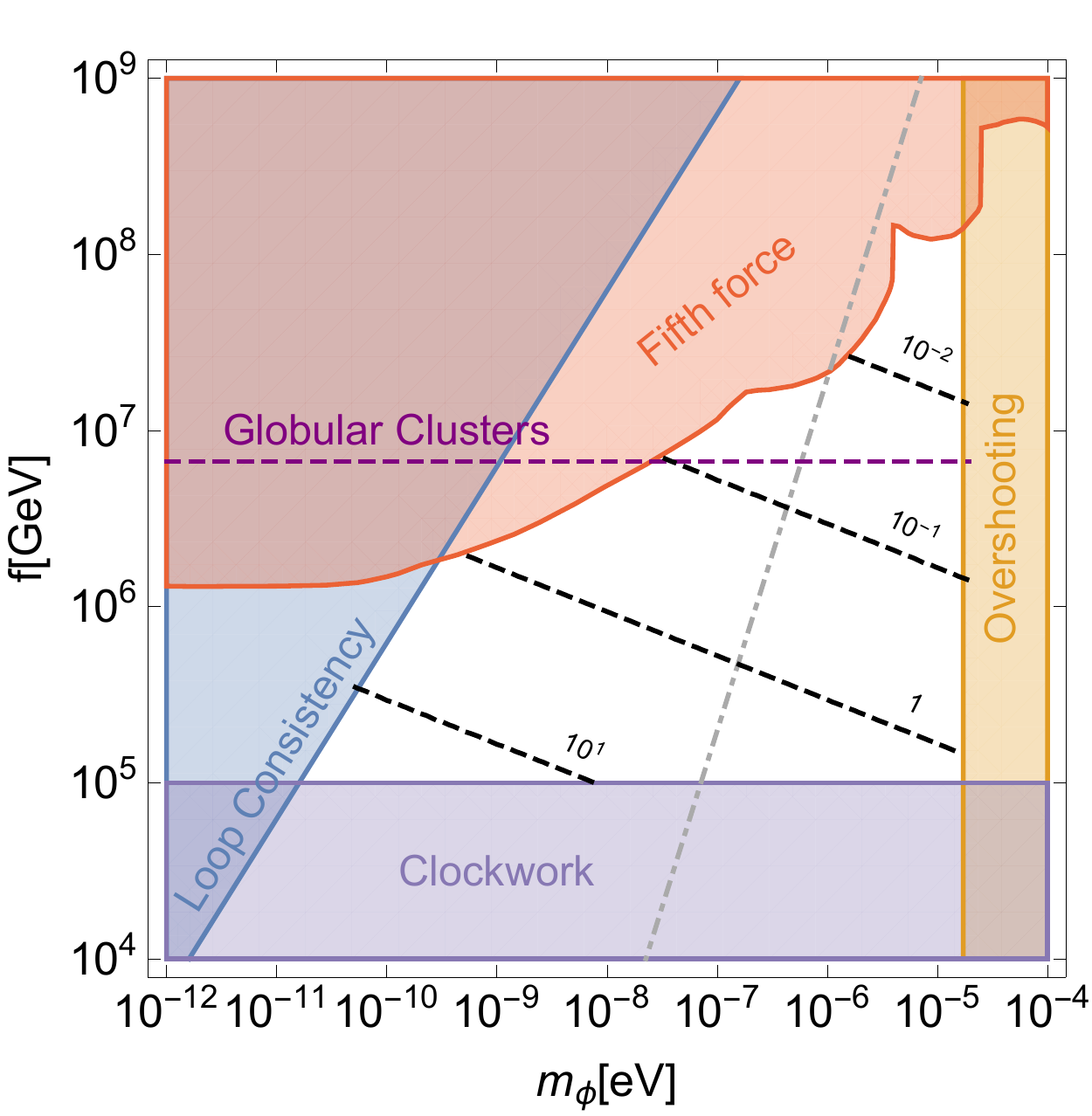}
\par\end{centering}
\protect\caption{The parameter space in which spontaneous baryogenesis happens during second roll of the relaxion. The orange shows a region where the relaxion does not get trapped by the back-reaction potential. The quantum corrections spoil the backreaction potential in the blue region. The region that leads to rapid star cooling in globular clusters is below the purple dashed line, however, this bound can be relaxed by a suitable choice of current. The purple region shows the clockwork requirement  $M\lesssim f$  for $M \gtrsim 100$ TeV. Finally, the fifth force constraints form mixing with the Higgs particle rule out region in red. We have added contours of  the value of $\tan(\phi_0/f)$ required   to obtain the correct relic density as per Ref.~\cite{Banerjee:2018xmn}. Finally, the grey line indicates the region where $\fw = \mpl$.\label{fig:paramspace}}
\end{figure}

\paragraph{Relaxion Constraints:} In order to realise the relaxion mechanism successfully, the loop consistency condition in Eq.~\eqref{eq:quantcorr} requires $\Lc^4 < 16\pi^2 v^4$, which translates into
\begin{equation}
    \frac{\pi^2}{810} \eta^2g_*^3 \frac{\TSPH^6}{ v^4 \mpl^2} ~<~ \frac{m_\phi^2}{f^2}~.
\end{equation}
In addition the slow roll requirement in Eq.~\eqref{eq:slowroll} provides an upper a bound on $m_\phi$:
\begin{equation}
    m_\phi ~<~ 5 \sqrt{\frac{g_* \Tc^4}{90 \mpl^2}} ~\sim~ 3.8\times 10^{-5}\;\mathrm{eV}\left(\frac{\Tc}{\TSPH}\right)^2.
\end{equation}
One can verify that this is the only bound that depends on $\fc = \Tc/\sqrt{4\pi}$. Therefore, changing $\fc$ effects only the upper bound on $m_\phi$.  Finally for a successful  clockwork implementation of the relaxion mechanism, it is essential that the explicit/anomalous breaking at the first and last site are much smaller than the decay constant which implies for the cutoff, $M\lesssim f$.   All these constraints are shown in our master plot in Fig.~\ref{fig:sbg}.

\paragraph{Cosmological constraints:} The Universe must reheat to sufficient temperature in order to activate sphalerons, $\Tr > \TSPH$. Furthermore, the sphalerons must decouple before the relaxion stops again, $\TSPH > \Tc$,  otherwise   CPT is restored while sphalerons are active and they erase any net baryon number.
This is satisfied by our aforementioned choice of $\fc$. The combination of these two conditions gives the required second period of  relaxion roll. Finally, Ref~\cite{DeSimone:2016ofp} shows that the back-reaction of $J^\mu$ on $\dot{\phi}$ is sufficiently small if $f > \TSPH \sim 130$ GeV.

\paragraph{Experimental constraints:} The coupling of $\phi$ to SM fields must not be detectable. 
There are two types of constraint: the first type arises from the mixing of $\phi$ with the Higgs boson which results in an emergent fifth force for the relaxion mass range relevant here.  This effect does not depend on the choice of $J^\mu$, so it can be treated independently. 
The mixing angle between the relaxion and the Higgs comes from the back-reaction potential in Eq.~(\ref{brstr}) and is of order
\begin{equation}
    \sin \theta ~\sim~ \frac{\Lc^4}{\fw v m_h^2}~.
\end{equation}
The constraints on this angle as function of $m_\phi$ have been presented in Ref.~\cite{Flacke:2016szy} and are reinterpreted in our master plot as the red exclusion area.  


The second type of constraint is from the coupling of the relaxion to SM particles via the operator in Eq.~\eqref{eq:term}. In our region of interest, the most important bounds on the coupling  of the relaxion to electrons, photons and nucleons arise from the fact that such a coupling  allows for a  more efficient cooling of stars and supernovae (see for eg. Ref.~\cite{Raffelt:1996wa}).  While most of these bounds can be evaded  if $J_\mu$ does not contain first generation fermionic currents~\footnote{We are assuming that flavour mixing effects that reintroduce a coupling to the first generation, once ${\cal O}_1$ is rotated to the mass basis, are small.} (recall that a coupling to $G \tilde{G}$ must be absent in any case as discussed in Sec.~\ref{sec:spbar}), a coupling to   photons would still be induced,
\begin{equation}
    E\frac{\alpha}{8\pi}\frac{\phi}{f} F_{\mu\nu}{\tilde F}^{\mu\nu}~
\end{equation}
where $E$ is the electromagnetic anomaly coefficient for the relaxion shift symmetry and we have ignored terms with a further $m_\phi^2/m_f^2$ suppression ($m_f$ being the mass of the fermions in the loop).  The best experimental constraint on the operator comes from bounds on the rate of cooling of globular cluster stars, $f/E > 2\times 10^7$ GeV~\cite{Raffelt:1996wa}. In our master plot we show this bound for $E=1$ as a purple-dashed  line, which will to currents such as $J^\mu = \bar{\mu}_R \gamma^\mu \mu_R$ or $J^\mu = \bar{t}_R \gamma^\mu t_R - \bar{b}_R\gamma^\mu b_R$.  

Clever choices for the global charges can, however, suppress or even lead to a vanishing $E$. Consider for example
\begin{align}
J_1^\mu &~=~ \bar{t}_R \gamma^\mu t_R - \bar{c}_R\gamma^\mu c_R~, \\
J_2^\mu &~=~ \bar{L}_2 \gamma^\mu L_2 + \bar{\mu}_R\gamma^\mu \mu_R~.
\end{align}
where $L_2$ is the second generation lepton doublet.  For both $J_1^\mu$ and $J_2^\mu$, $E=0$ and the bound can be entirely evaded.  For $J_1^\mu$, it might seem that  any baryon number generation would cancel between the two terms but, as shown in Sec.~\ref{sec:spbar}, the mass difference between the top and charm can become important once one takes into account finite temperature effects. Indeed an analysis following that in Sec.~\ref{sec:spbar} readily leads to factor proportional to
$\Lambda_t-\Lambda_c$ in the baryon asymmetry corresponding to this  current, which is not particularly suppressed.

 Note that the above charge assignments have no colour anomaly and are thus consistent  with the requirement  of no strong CP phase being  generated when the relaxion stops (see Sec.~\ref{sec:spbar}).   Finally we should remark that for a full flavour model, the relaxion can induce flavour violation once the operator in Eq.~\eqref{eq:term} is rotated to the mass basis~\cite{Davidi:2018sii}.

\paragraph{Dark Matter:}
Ref \cite{Banerjee:2018xmn} points out that the relaxion oscillations can explain the observed dark matter abundance. At any point in the plane of Fig.~\ref{fig:paramspace}, the expected dark matter abundance can be computed up to the square of the factor $\tan(\phi_0/f)$ (see Eq.~\eqref{eq:misalignment} and Eq.~\eqref{eq:dm}).   We show in  Fig.~\ref{fig:paramspace} lines indicating  the value of $\tan(\phi_0/f)$ required to match the observed dark matter abundance. Values of $\tan(\phi_0/f)$ much larger or smaller than unity are an indication of the level of tuning required to obtain the correct relic density (see the discussion below Eq.~\eqref{eq:vprime}). This is especially  true if $\tan(\phi_0/f)\ll1$ as the universe will overclose without this tuning whereas in the regions with $\tan(\phi_0/f)\gg1$, relaxion oscialltions would only account or a small fraction of the observed dark matter density without tuning. Remarkably, we find that in the small allowed region obtained after imposing all the theoretical and experimental constraints discussed here,   the required tuning to obtain observed relic density ranges from none at all to a maximum of ~$\sim 100$.

Finally, the relaxion DM should be stable and not decay to photons. We have checked that, even in the example with the least suppressed  couplings discussed above, the relaxion decay time to two photons is much longer than the age of the Universe: $\tau \gg H_0^{-1}$.

\section{Discussion and Conclusion}

We have shown that Spontaneous Baryogenesis is an integral feature of relaxion models with a high reheat temperature, which induces a second stage of rolling, and that it can generate the SM baryon asymmetry we observe today with almost no adjustment. Indeed the only additional ingredient is a single operator coupling the relaxion to matter currents, $\partial_\mu \phi J^\mu$, which arises automatically if some SM fields are charged under the U(1) of which the relaxion is a pseudo-nambu goldstone boson.  Remarkably one encounters (to adapt a phrase from \cite{Banerjee:2018xmn}) a {\it double} ``relaxion-miracle'', because the baryogenesis mechanism operates most effectively, and with least tuning, exactly where the relaxion oscillations are a viable dark matter candidate. \\

\noindent
{\it Acknowledgements} This work was supported by IPPP Grant ST/P001246/1. 

\pagebreak

\bibliography{citations}
\nocite{*}
\bibliographystyle{unsrt}

\end{document}